\documentclass[12pt]{iopart}

 \usepackage{graphicx}
 \usepackage{amssymb}
 \usepackage{bm}      
 \usepackage{verbatim}%
 \usepackage{epstopdf}
 \usepackage[light,timestamp,draft]{draftcopy} 
 \usepackage{iopams}  

\begin{document}

\title[]{
Molecular dynamics simulation of Ga penetration along Al grain boundaries under a constant strain rate condition
}

\author{Ho-Seok Nam}
\address
{
School of Advanced Materials Engineering, Kookmin University, Seoul 136-702, Korea
}
\ead{hsnam@kookmin.ac.kr}
\author{David J. Srolovitz}

\address
{
Institute of High Performance Computing, Singapore 138632, Singapore
}

\begin{abstract}
While diverse fracture characteristics have been observed in liquid metal embrittlement (LME) depending on the solid-liquid metal pairs, the penetration of nanometer-thick liquid metal films along the grain boundary has been identified as one of the key mechanisms for embrittlement in many classical LME systems, such as Al-Ga, Cu-Bi and Ni-Bi.  For example, liquid Ga quickly penetrates deep into grain boundaries in Al, leading to intergranular fracture under very small stresses.  We report on a series of molecular dynamics simulations of liquid Ga in contact with an Al bicrystal under a constant strain rate.  We identify the grain boundary dislocations that are nucleated at the grain boundary groove tip and climb down along the grain boundary during Ga penetration and characterize their atomic structures based on topological method.
\end{abstract}

\vspace{2pc}


\section{Introduction}
\label{intro}

One of the most interesting environmental failure phenomena in polycrystalline metals is the degradation of the mechanical properties of ductile metals in the presence of an aggressive, wetting liquid metal and an external stress, which is called liquid metal embrittlement (LME).  This phenomenon presents a challenge in a wide range of material processing scenarios (including welding/brazing/soldering, galvanizing, heat treatment, hot working, etc.) and is particularly important in nuclear reactors in which liquid metals are used as coolants or as spallation targets.

A wide range of solid-liquid metal pairs shows LME.  Although there are some common features of LME systems, several kinds of fracture characteristics have been observed in LME depending on the solid-liquid metal couple~\cite{Joseph:LMEreview}.  Among those, the penetration of nanometer-thick liquid metal films along the grain boundary has been identified as one of the key mechanisms for embrittlement in many classical LME systems (e.g.,  Al-Ga, Cu-Bi and Ni-Bi).  The Al-Ga couple is probably the most widely studied among the systems known for the formation of penetrating intergranular films.  It exhibits strikingly rapid Ga penetration along grain boundaries, resulting in a very rapid, dramatic loss of Al ductility.  Many experimental studies show that liquid Ga penetrates into Al grain boundaries at a remarkable rate ($\sim$several $\mu$m/s at room temperature), leading to a distinct channel morphology~\cite{Hugo:AlGaTEM,Kozlova:AlGaSEM,Pereiro-Lopez:AlGaPRL2005,Pereiro-Lopez:AlGaPolycrystal2004,Ludwig:AlGaBicrystal2005,Pereiro-Lopez:AlGaBicrystal2006}.  The penetration of liquid Ga along the grain boundaries produces wetting layers with thicknesses reported to range from a few monolayers~\cite{Hugo:AlGaTEM} to several hundred nanometers~\cite{Pereiro-Lopez:AlGaPRL2005,Pereiro-Lopez:AlGaPolycrystal2004,Ludwig:AlGaBicrystal2005}, even in the absence of an applied load.  Interestingly, the rate of propagation of such liquid layers is strongly influenced by even very small stresses~\cite{Pereiro-Lopez:AlGaPolycrystal2004,Ludwig:AlGaBicrystal2005,Pereiro-Lopez:AlGaBicrystal2006}.

Several models have been proposed to explain the driving forces and atomic mechanisms by which the liquid phase penetrates quickly along grain boundaries~\cite{Bokstein:DiffusionDisolution,Glickman:DCM,Rabkin:CoherencyStresses}. However, many fundamental questions remain unsettled for the Al-Ga couple.  For example, none of these approaches successfully explains the effects of stress on liquid film penetration.  Recently, Nam and Srolovitz~\cite{HoseokNam:PRL} examined LME through the lens of molecular dynamics (MD) simulations of an Al bicrystal in contact with liquid Ga and observed how Ga penetrates along the grain boundaries with and without an applied stress.  The simulation study proposed a mechanism of LME and clarified how it is affected by applied stresses.  The interplay of stress and  Ga penetration leads to the nucleation of a train of dislocations on the grain boundary below the liquid groove root which climbs down the grain boundary at a nearly constant rate.

In this study, we carry out MD simulations of stress-facilitated Ga penetration under a constant strain rate condition.
We further investigate on the characteristics of the grain boundary dislocations that are consistently observed in our series of studies~\cite{HoseokNam:PRL,HoseokNam:PRB,HoseokNam:Acta}:  We analyze atomic structures of the grain boundary dislocations in more detail, based on the topological approach by Hirth and Pond~\cite{Hirth:Topological}.  Our results show that the characteristics of climbing grain boundary dislocations would depend intrinsically on the grain boundary type/structure.

\section{Simulations Procedure}
\label{sec:1}

\begin{figure*}[!tbp]
\includegraphics[width=0.47\textwidth]{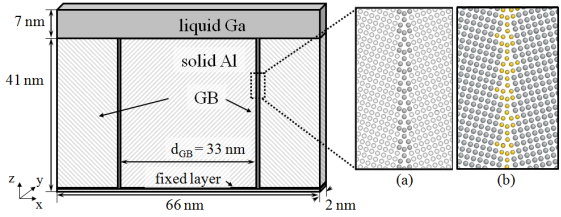}
\caption
{ \label{fig:Geometry}
The simulation cell contains two Al grains ($\sim 320,000$ atoms) in contact with liquid Ga ($\sim 40,000$ atoms).  Panels (a) and (b) show the atomic structures of (a) the $\Sigma 5~ 36.9 \,^{\circ} (301) /[010]$ symmetric tilt boundary and (b) the $\Sigma 17~ 28.1 \,^{\circ} (401) /[010]$ symmetric tilt boundary, respectively.
}
\end{figure*}

All of the simulations were performed on a 3-dimensional Al bicrystal sample in contact with liquid Ga, as shown schematically in Fig.~\ref{fig:Geometry}.  Liquid Ga was initially presaturated with Al (the solubility limit as a function of temperature for these interatomic potentials was determined in Ref.~\cite{HoseokNam:PRB}).  The atomic interactions in the Al-Ga system were described by semiempirical embedded-atom method (EAM) potentials developed in the previous study~\cite{HoseokNam:PRB}.  We imposed periodic boundary conditions in the $x$- and $y$-directions, fixed several atomic layers at the bottom of the bicrystal (to prevent grain rotation so that it mimics the surface region of a macroscopic sample) and left the top surface free (i.e., there is a vacuum above the liquid)~\cite{HoseokNam:PRB}.  (For more details, see Ref.~\cite{HoseokNam:PRB}.)

In order to investigate the stress-facilitated Ga penetration, the MD simulations were performed under a constant strain rate condition ($NVT$ ensemble) of $\epsilon_{xx}=\epsilon_{0}+\dot{\epsilon}_{xx}$, where the strain rate $\dot{\epsilon}_{xx}$ was controlled by stretching the simulation cell every 1 ps at a constant rate in the $x$-direction ($\epsilon_{0}$ is a initial strain).  We performed the simulations at a constant strain rate rather than fixed grip or fixed load conditions in order to complement the artificial grain size effect.  (More discussion is provided in the next section.)  The simulations were performed at 600 K with a strain rate condition of $\epsilon_{0}=0.01$ and $\dot{\epsilon}_{xx}=1 \times 10^5 s^{-1}$.

A couple of different grain boundary types were examined including $\Sigma 5~ 36.9 \,^{\circ} (301)/[010]$ symmetric tilt boundaries, $\Sigma 17~ 28.1 \,^{\circ} (401)/[010]$ symmetric tilt boundaries, and low angle tilt boundaries.  Both $\Sigma 5$ and $\Sigma 17$ tilt grain boundaries have relatively high grain boundary energies relative to the low angle tilt grain boundaries.

The MD simulations were performed using the Large-scale Atomic/Molecular Massively Parallel Simulator  (LAMMPS) code~\cite{Plimpton:ParallelAlgorithm,LAMMPS:homepage}. The equations of motion were integrated using the velocity Verlet method.  Total simulation time was at least 60 ns ($\sim 2 \times 10^7 \Delta t$, where the time step $\Delta t=$2.5 fs).

\section{Result and Discussion}
\label{sec:2}

\begin{figure*}[!tbp]
\includegraphics[width=0.67\textwidth]{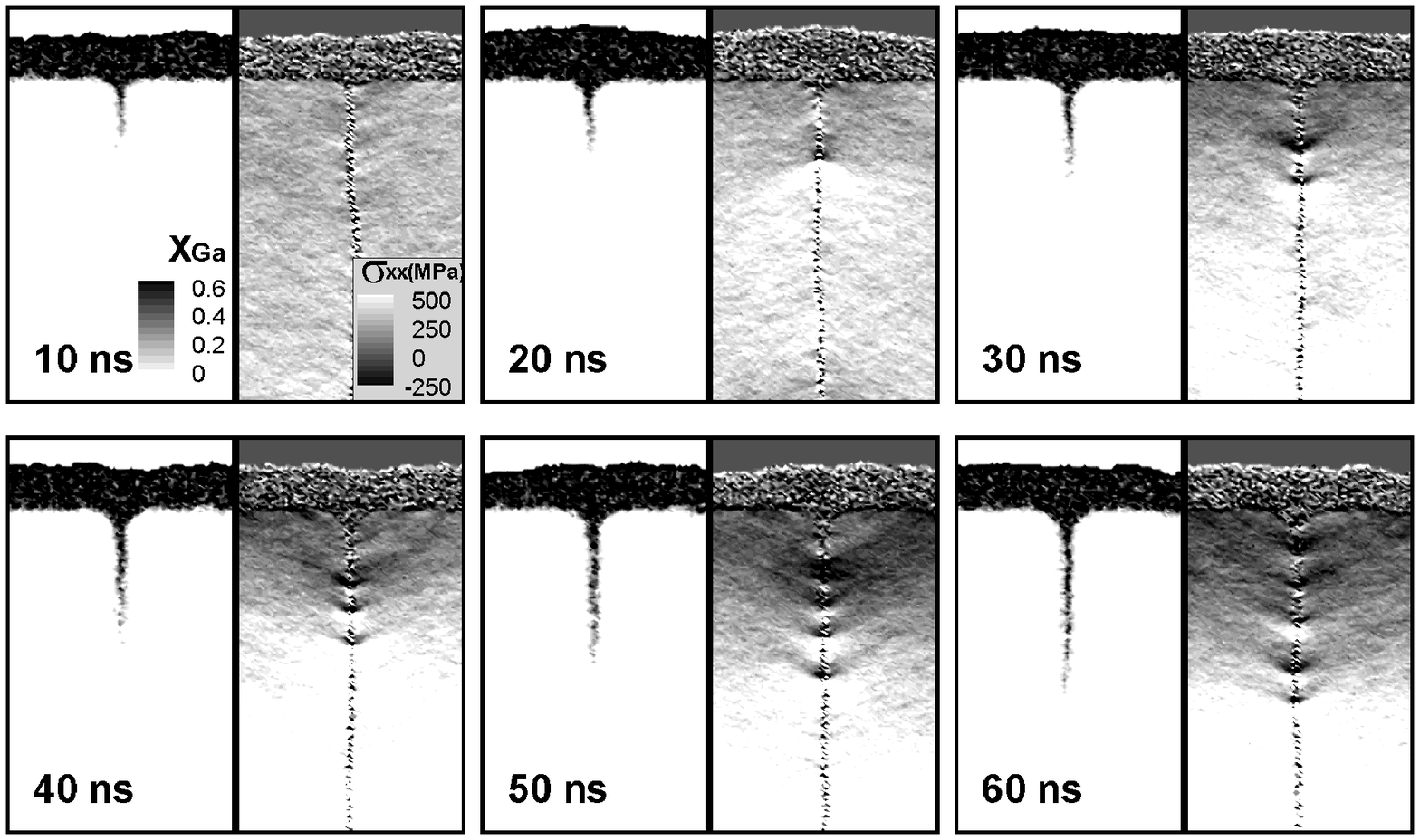}
\includegraphics[width=0.32\textwidth]{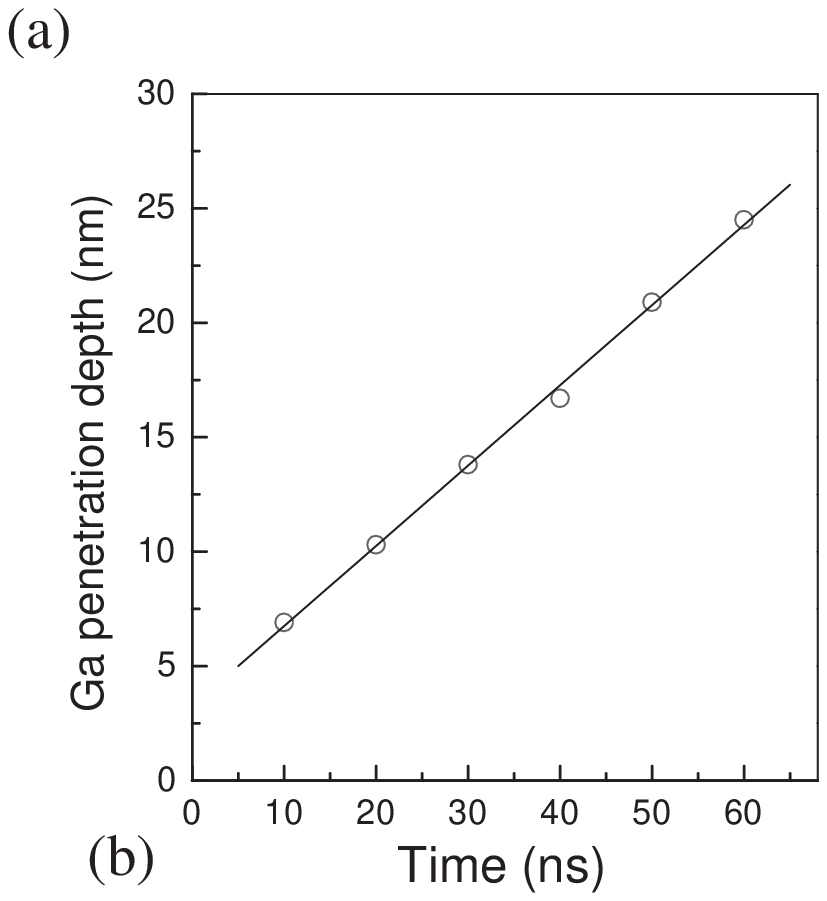}
\caption
{ \label{fig:StrainRate}
(a) Ga concentration (left panel) and stress $\sigma_{xx}$ (right panel) profiles at $10$ to $60$ ns for simulations at 600 K at a constant strain rates of $\sim 1 \times 10^5 s^{-1}$.  The Ga penetration depth vs. time is shown in (b).
}
\end{figure*}

We performed a series of MD simulations of an Al bicrystal in contact with liquid Ga and examined the initial stage of Ga penetration along the Al grain boundaries at 600 K under a constant strain rate condition of $\epsilon_{0}=0.01$ and $\dot{\epsilon}_{xx}=1 \times 10^5 s^{-1}$.  The imposed strain rate condition corresponds to change of strain from $1.0\%$ to $1.6\%$ for our 60 ns simulation time.  Figure~\ref{fig:StrainRate}(a) shows the Ga concentration and stress profiles at $10$ to $60$ ns for simulations of the $\Sigma 5~ 36.9 \,^{\circ} (301)/[010]$ symmetric tilt boundary.  In Fig.~\ref{fig:StrainRate}(b), the Ga penetration depth vs. time is shown, where the penetration depth is arbitrarily defined as the distance from the surface (in the $z$-direction) where the Ga content is equivalent to 0.3 monolayers (a typical Ga content profile was shown in Ref.~\cite{HoseokNam:PRB}).  Our MD simulations under a constant strain rate condition show similar penetration behavior of liquid Ga along the Al grain boundary as under fixed grip conditions~\cite{HoseokNam:PRB}.  Application of a constant strain rate significantly promotes liquid metal penetration by forming a set of climbing grain boundary dislocations, resulting in fixed rate penetration mode.  However, we could observe nucleation and climbing of more grain boundary dislocations than in the previous MD simulations under fixed grip conditions~\cite{HoseokNam:PRB}.  Interestingly, grain boundary dislocations are generated continuously all over the simulation time.

In fact, implement of appropriate stress condition is a central issue in our simulation of the stress-facilitated Ga penetration.  For example, difference between constant stress ($NPT$) and constant strain ($NVT$) ensemble simulations can be quite significant in this situation.  If we implemented a constant \emph{stress} algorithm (fixed load condition), the grain boundary traction within the simulation cell would never drop all over the simulation time (with no grain size effect).  However, the grain boundary would pull apart as liquid Ga penetrated only a few tens of nanometers.  This is clearly unphysical, since our simulation cell is meant only to mimic a very thin layer near the surface of a very thick polycrystalline sample, where most of the load is carried by the solid, far from the surface.  On the other hand, if we performed simulations under a constant \emph{strain} (fixed grip) condition, the applied stress effect would fade out in relatively short simulation time since the initial stress within the fixed system would be quickly relaxed by even very small amount of Ga penetration along the grain boundary.  This unintended grain size effect is unavoidable because the inter-distance of the grain boundaries (under the current simulation cell size and periodic boundary condition) is too small compared to macroscopic size.  In order to reflect real grain size sample, we might carry out full-size multiscale simulations, e.g, where the displacement field for a macroscopic sample is calculated by using a continuum approach and the atomic process that occurs in the embedded region (where the grain boundary intersects the solid-liquid interface) is simulated by molecular dynamics.  While this could, in principle, be done, the computational cost and complexity is too expensive to implement.  Instead, Namilae \emph{et al}~\cite{Namilae:MSMSE} applied artificial \emph{crack tip opening loads} on Al bicrystals (with a pre-existing notch) in the presence of liquid Ga, where regions adjacent to the wedge were subject to a constant displacement rate.  In this way, they studied the effect of Ga on the crack opening of Al bicrystals using atomistic simulations.  Here, we performed MD simulations under \emph{constant strain rate} conditions as a compromise.

Clearly, the grain boundary traction in real sample would be relieved as Ga penetrates along the grain boundary.  However, the rate of stress relieving near grain boundaries in an macroscopic polycrystalline sample would be much slower (by orders of magnitude off) than we could expect in a constant strain simulation of nanometer .  In this respect, application of appropriative constant strain rate starting from a finite strain value is expected to compensate the size discrepancy in a constant strain simulation.  In actual situation, the strain rate of embedded region would be a time dependent variable coupled with both stress redistribution by Ga penetration at the embedded grain boundary region and accompanying relaxation within the surrounding large grains far from the grain boundary.  However, based on the linear Ga penetration kinetics of our previous simulation, we expect that the strain rate near the grain boundary would be approximately constant within the initial stage of Ga penetration and treated as a magic number parameter for simplicity.  With the application of appropriate constant strain rates, the nucleation of grain boundary dislocations was dramatically increased (up to more than 5 dislocations within our 60 ns simulation time), while the stresses within the system remained steady by balancing the displacements introduced by Ga penetration into the grain boundary and the increase of the simulation cell size.  

\begin{figure}[!tbp]

\includegraphics[width=0.85\textwidth]{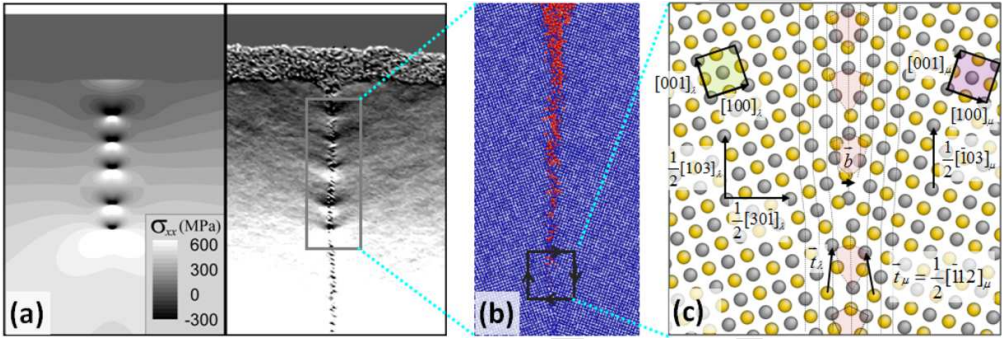}

\caption
{ \label{fig:DislocationCore}
(Color) (a) Comparison of the MD simulation (right) and linear elastic stress field (left)
(b) Atomic scale image of Ga penetration along $\Sigma 5$ (301)/[010] symmetric tilt boundary in an Al bicrystal. The atoms shown in blue represent Al atoms and those in red are Ga.
(c) Enlarged view of dislocation core region (Burgers circuit). Atomic structure was quenched in order to remove thermal noise. The Grey and yellow circles represents rows of atoms with positions in alternating (020) planes.
}
\end{figure}

Next, in order to characterize the grain boundary dislocations, we examined the atomic configuration near the stress concentration patterns.  Figure~\ref{fig:DislocationCore}(b) shows the atomic scale image of Ga penetration along $\Sigma 5$ symmetric tilt boundary captured from the box region indicated in Fig.~\ref{fig:DislocationCore}(a).  The atoms shown in blue represent Al atoms and those in red are Ga.  After quenching the atomic structure in order to remove thermal noise, atomic scale image was further enlarged as shown in Fig.~\ref{fig:DislocationCore}(c).  Examination of the atomic structure of the grain boundary shows the existence of an interfacial dislocation at the location of the center of this stress pattern.  We could accurately determine the Burgers vector from atomic scale images of the dislocation by applying the topological theory of linear defects at interfaces suggested by Hirth and Pond~\cite{Hirth:Topological}.  The Burgers circuit was indicated in Fig.~\ref{fig:DislocationCore}(b) and the analysis approach is shown in Fig.~\ref{fig:DislocationCore}(c) for the $\Sigma 5$ symmetric tilt boundary examined in this study.  Using this approach, we found that this LME dislocation has Burgers vector $\vec{b}$ given by the following relation:
\begin{equation}
\label{eq:DSCvector}
\vec{b}=\vec{t}_{\lambda} - \vec{t}_{\mu} = \frac{1}{10}[30\bar{1}]_{\lambda},
\end{equation}
where $\vec{t}_{\lambda}$ and $\vec{t}_{\mu}$ are step vectors in the crystal systems of the grain $\lambda$ on the left and the grain $\mu$ on the right, respectively.

The Burgers vector is indeed the displacement shift complete (DSC) vector of $\vec{b} = \frac{1}{10}[\bar{3}10]_{\lambda}$ in the crystal system of the grain $\lambda$.  The component of Burgers vector normal to grain boundary is now $b_n = |\vec{b}| = \frac{a_0}{\sqrt{10}} \simeq$ 1.28 \AA, where $a_0$ is the cubic lattice parameter.  We confirm this measurement by analytically calculating the stress field associated with such dislocations~\cite{Head:EdgeDislStress} using the measured bicrystal elastic properties and comparing it with the stress field determined in the simulation, as shown in Fig.~\ref{fig:DislocationCore}(a).  The excellent correspondence confirms that the normal component of the Burgers vector $b_n$ is $\sim 1.28$ \AA.

A series of [010] symmetric tilt boundaries was examined in the present simulations: $\Sigma 5~[010]$ symmetric tilt boundaries, $\Sigma 17~[010]$ symmetric tilt boundaries, and low angle tilt boundaries.  We found that the Ga penetration behavior is sensitive to grain boundary type and structure: no grain boundary wetting was observed within our 60 ns simulations for the low angle ($\sim 10 \,^{\circ}$) grain boundaries, while the $\Sigma 5~ [010]$ and $\Sigma 17~ [010]$ tilt boundaries showed remarkable Ga penetration rates with formation and climbing of LME dislocations.  We also analyzed atomic structure of LME dislocation for the $\Sigma 17~ [010]$ tilt boundaries in the same procedure as described above and found that the Burgers vector is also the DSC vector with the normal component of the Burgers vector $b_n = |\vec{b}| = \frac{a_0}{\sqrt{17}} \simeq$ 0.98 \AA.  The Burgers vector analysis implies that characteristics of LME dislocation should be dependent on grain boundary crystallography and this explains why Ga penetration kinetics is sensitive to grain boundary type/structure.

Similar to the previous reports~\cite{HoseokNam:PRL,HoseokNam:PRB,HoseokNam:Acta}, our simulations demonstrate that application of a constant strain rate significantly promotes liquid metal penetration along grain boundaries, resulting in a change from a diffusive to fixed rate penetration mode.  First, Ga diffuses down the grain boundary in Al below the liquid groove root and causes stresses large enough to nucleate a dislocation in the grain boundary.  The first dislocation ``climbs'' down by stress-enhanced Ga hoping across the dislocation core, leaving a tail of Ga behind.  This Ga hopping leads to a constant dislocation climb rate that is applied stress-independent. Once the dislocation moves far enough from the groove root, another dislocation is nucleated.  It too climbs down the grain boundary at the same rate, resulting in a uniform spacing of climbing dislocations.  With Ga at the grain boundary, applied strains enhance the grain boundary opening and in turn more Ga is inserted from the liquid groove into the grain boundary to relieve the residual stress (i.e., Ga layer thickening process).  The Ga penetration rate mirrors the dislocation climb rate and hence is time independent. Here, the applied strain rate provides stress field to aid the nucleation of dislocations at the grain boundary and keep the grain boundary open to allow sufficient Ga transport enough to move with the dislocation.

\section{Summary}
\label{sec:3}

We performed MD simulations of an Al bicrystal in contact with liquid Ga and examined the penetration of Ga along the Al grain boundaries under a constant strain rate condition.  Consistent with the previous reports~\cite{HoseokNam:PRL,HoseokNam:PRB,HoseokNam:Acta}, we could observe that LME dislocations are nucleated and climb down along grain boundary, resulting in fixed rate penetration mode.  With the application of a constant strain rate, the nucleation of grain boundary dislocations was dramatically increased (up to more than 5 dislocations within our 60 ns simulation time), while the stress within the system remained steady by balancing the displacements introduced by Ga penetration into the grain boundary and the increase of the simulation cell size.  We also analyzed the grain boundary dislocations based on topological approach and characterized it in the framework of interfacial structure.  Our analysis on several different grain boundaries show that the characteristics of climbing grain boundary dislocations depend on the grain boundary type/structure and this implies that Ga penetration rate could be sensitive to grain boundary type/structure due to the different topological properties of grain boundary dislocation.


\ack
H.-S. Nam gratefully acknowledge the support of the faculty research program 2008 of Kookmin University, the support of the Korea Research Foundation Grant funded by the Korean Government (MOEHRD, Basic Research Promotion Fund) (KRF-2008-331-D00269), and the Priority Research Centers Program through the National Research Foundation of Korea (NRF) funded by the Ministry of Education, Science and Technology (2009-0093814).



\section*{References}

\begin{thebibliography}{18}

\bibitem{Joseph:LMEreview}
B. Joseph, M. Picat, and F. Barbier,
              Eur. Phys. J. AP {\bf 37}, 875 (1999).

\bibitem{Hugo:AlGaTEM}
R.~C. Hugo and R.~G. Hoagland,
              Scripta Mater.   {\bf 38}, 523  (1998);
              {\it ibid.}      {\bf 41}, 1341 (1999);
              Acta Mater.      {\bf 48}, 1949 (2000).

\bibitem{Kozlova:AlGaSEM}
O. Kozlova {\textit et al.},
              Def. Diff. Forum {\bf 237}, 751 (2005);
O. Kozlova and A. Rodin,
              {\it ibid.}      {\bf 249}, 231 (2006).

\bibitem{Pereiro-Lopez:AlGaPRL2005}
E. Pereiro-L\'{o}pez, W. Ludwig, D. Bellet, P. Cloetens, and C. Lemaignan,
              Phys. Rev. Lett. {\bf 95}, 215501 (2005).

\bibitem{Pereiro-Lopez:AlGaPolycrystal2004}
E. Pereiro-L\'{o}pez, W. Ludwig, and D. Bellet,
              Acta Mater.      {\bf 52}, 321 (2004);

\bibitem{Ludwig:AlGaBicrystal2005}
W. Ludwig, E. Pereiro-L\'{o}pez, and D. Bellet,
              Acta Mater.      {\bf 53}, 151 (2005);

\bibitem{Pereiro-Lopez:AlGaBicrystal2006}
E. Pereiro-L\'{o}pez, W. Ludwig, D. Bellet, and P. Lemaignan,
              Acta Mater.      {\bf 54}, 4307 (2006).

\bibitem{Bokstein:DiffusionDisolution}
B.~S. Bokstein, L.~M. Klinger, and I.~V. Apikhtina,
              Mat. Sci. Eng. A {\bf 203}, 373 (1995).

\bibitem{Glickman:DCM}
W.~M. Robertson,
              Trans. Metall. Soc. AIME {\bf 236}, 1478 (1966);
E.~E. Glickman,
              Z. Metallkd      {\bf 96}, 1204 (2005).

\bibitem{Rabkin:CoherencyStresses}
E. Rabkin,
              Scripta Mater.   {\bf 39}, 685 (1998).

\bibitem{HoseokNam:PRL}
H.-S. Nam and D.~J. Srolovitz,
              Phys. Rev. Lett. {\bf 99}, 25501 (2007).

\bibitem{HoseokNam:PRB}
H.-S. Nam and D.~J. Srolovitz,
              Phys. Rev. B     {\bf 76}, 184114 (2007).

\bibitem{HoseokNam:Acta}
H.-S. Nam and D.~J. Srolovitz,
              Acta mater.      {\bf 57}, 1546 (2009).

\bibitem{Hirth:Topological}
J.~P. Hirth and R.~C. Pond,
              Acta mater. {\bf 44}, 4749 (1996).

\bibitem{Plimpton:ParallelAlgorithm}
S. Plimpton,
              J. Comput. Phys. {\bf 117}, 1 (1995).

\bibitem{LAMMPS:homepage}
http://lammps.sandia.gov

\bibitem{Namilae:MSMSE}
S. Namilae, B. Radhakrishnan and J.~R. Morris,
              Modelling Simul. Mater. Sci. Eng. {\bf 16}, 075001 (2008).

\bibitem{Head:EdgeDislStress}
A.~K. Head,
              Proc. Phys. Soc. London. {\bf B66}, 793 (1953).

\end{thebibliography}
\end{document}